%% file: MAIN.tex
\documentclass[sigconf]{acmart}

\usepackage{colortbl}
\usepackage{floatflt}
\usepackage{wrapfig}

\definecolor{Gray1}{gray}{0.9}
\definecolor{LightGreen}{rgb}{1,1,0.92}
\definecolor{LightCyan}{rgb}{0.95,1,1}

\usepackage{algorithmic}
\usepackage{graphicx}
\usepackage{hyperref}
\usepackage{textcomp}
\usepackage{alltt}
\usepackage{array} 

\usepackage{xcolor}
\newcommand{\framedtext}[1]{%
  \par\medskip\noindent
  \begingroup
    \setlength{\fboxsep}{4pt}% inner padding
    \setlength{\fboxrule}{0.5pt}% border thickness
    \colorlet{cbg}{gray!10}%
    \colorlet{cbd}{gray!70!black}%
    \fcolorbox{cbd}{cbg}{%
      \parbox{\dimexpr\linewidth-2\fboxsep-2\fboxrule\relax}{%
        \sffamily\small #1%
      }%
    }%
  \endgroup
  \par\medskip
}

\newcommand\blfootnote[1]{%
  \begingroup
  \renewcommand\thefootnote{}\footnote{#1}%
  \addtocounter{footnote}{-1}%
  \endgroup
}

\AtBeginDocument{%
  }

\setcopyright{acmlicensed}
\copyrightyear{2025}
\acmYear{2025}
\acmDOI{XXXXXXX.XXXXXXX}
\acmConference[INDIS 2025]{International Workshop on Innovating the Network for Data-Intensive Science in conjunction with SC25}{November 16--21, 2025}{St. Louis, MO, USA}
\acmISBN{978-1-4503-XXXX-X/2018/06}

\begin{document}

\title{From Edge to HPC: Investigating Cross-Facility Data Streaming Architectures}


\author{Anjus George}
\affiliation{%
  \institution{National Center for Computational Sciences, Oak Ridge National Laboratory}
  \city{Oak Ridge, TN}
  \country{USA}}
\email{georgea@ornl.gov}

\author{Michael Brim}
\affiliation{%
  \institution{National Center for Computational Sciences, Oak Ridge National Laboratory}
  \city{Oak Ridge, TN}
  \country{USA}}
\email{brimmj@ornl.gov}

\author{Christopher Zimmer}
\affiliation{%
  \institution{National Center for Computational Sciences, Oak Ridge National Laboratory}
  \city{Oak Ridge, TN}
  \country{USA}}
\email{zimmercj@ornl.gov}

\author{David Rogers}
\affiliation{%
  \institution{National Center for Computational Sciences, Oak Ridge National Laboratory}
  \city{Oak Ridge, TN}
  \country{USA}}
\email{rogersdm@ornl.gov}

\author{Sarp Oral}
\affiliation{%
  \institution{National Center for Computational Sciences, Oak Ridge National Laboratory}
  \city{Oak Ridge, TN}
  \country{USA}}
\email{oralhs@ornl.gov}

\author{Zach Mayes}
\affiliation{%
  \institution{National Center for Computational Sciences, Oak Ridge National Laboratory}
  \city{Oak Ridge, TN}
  \country{USA}}
\email{mayeszb@ornl.gov}

\renewcommand{\shortauthors}{George et al.}

\begin{abstract}
In this paper, we investigate three cross-facility data streaming architectures, Direct Streaming (DTS), Proxied Streaming (PRS), and Managed Service Streaming (MSS). We examine their architectural variations in data flow paths and deployment feasibility, and detail their implementation using the Data Streaming to HPC (DS2HPC) architectural framework and the SciStream memory-to-memory streaming toolkit on the production-grade Advanced Computing Ecosystem (ACE) infrastructure at Oak Ridge Leadership Computing Facility (OLCF). We present a workflow-specific evaluation of these architectures using three synthetic workloads derived from the streaming characteristics of scientific workflows. Through simulated experiments, we measure streaming throughput, round-trip time, and overhead under work sharing, work sharing with feedback, and broadcast and gather messaging patterns commonly found in AI-HPC communication motifs. Our study shows that DTS offers a minimal-hop path, resulting in higher throughput and lower latency, whereas MSS provides greater deployment feasibility and scalability across multiple users but incurs significant overhead. PRS lies in between, offering a scalable architecture whose performance matches DTS in most cases.
\end{abstract}





\maketitle

\input{tex/introduction}

\input{tex/architecture}

\input{tex/framework}

\input{tex/deployment}

\input{tex/evaluation}

\input{tex/discussion}

\begin{acks}
We thank our colleagues Nick Schmitt, Ethan O’Dell, Steven Lu, Tyler Skluzacek, A.J. Ruckman, Paul Bryant, and Gustav Jansen at ORNL for their help in resolving technical challenges during the deployment of the streaming architectures and for providing valuable information and clarifications. We also thank the SciStream team at ANL, Flavio Castro and Rajkumar Kettimuthu, for helping us better understand SciStream’s capabilities. This research used resources of the Oak Ridge Leadership Computing Facility located at Oak Ridge National Laboratory, which is supported by the Office of Science of the Department of Energy under contract No. DE-AC05-00OR22725.
\end{acks}

%\bibliographystyle{ACM-Reference-Format}
%\bibliography{REFERENCES}


%\appendix

\end{document}

%% file: tex/introduction.tex
\section{Introduction}

The scientific computing landscape is being reshaped by the convergence of artificial intelligence (AI), foundation models, digital-twins, and exascale simulations \cite{aicoupledhpcworkflows, material_discovery, weather-forecasting, wes-aicoupled, biomolecular}. These advances have sparked new momentum for modern workflows that increasingly demand near real-time data analysis, experimental steering, and informed decision-making during experiment execution. To meet these demands, researchers are developing tightly coupled workflows that integrate experimental facilities with high-performance computing (HPC) systems, enabling on-the-fly analysis and timely feedback to guide experimental control. \blfootnote{This manuscript has been authored by UT-Battelle, LLC under Contract No. DE-AC05-00OR22725 with the U.S. Department of Energy. The United States Government retains and the publisher, by accepting the article for publication, acknowledges that the United States Government retains a non-exclusive, paid-up, irrevocable, world-wide license to publish or reproduce the published form of this manuscript, or allow others to do so, for United States Government purposes. The Department of Energy will provide public access to these results of federally sponsored research in accordance with the DOE Public Access Plan. (http://energy.gov/downloads/doe-public-access-plan).} 

Recent examples of such experimental–HPC workflows include the Linac Coherent Light Source (LCLS) workflow at SLAC National Accelerator Laboratory, which streams diffraction frames from the LCLS light source over ESnet \cite{esnet} directly to HPC systems at OLCF \cite{olcf-lcls-news}. There, AI models identify Bragg peaks and recommend parameter changes while the sample is still in the beam \cite{lcls}. At Argonne’s Advanced Photon Source (APS), an AI enabled workflow pushes 2 kHz ptychography data to embedded GPUs for edge inference, while HPC nodes retrain the model on the fly, enabling dose-efficient imaging and real-time experimental steering \cite{aps-edge-dl}. Another example is an edge-to-exascale workflow using Frontier \cite{frontier} at OLCF, where a Temporal Fusion Transformer (TFT) is employed at the Spallation Neutron Source (SNS) to predict 3D scattering patterns and adjust beam settings within minutes instead of hours \cite{olcf-neutron-edge}. 

The U.S. Department of Energy (DOE) Integrated Research Infrastructure (IRI) initiative aims to accelerate the time to insight from such tightly coupled experimental-HPC workflows \cite{iri-program}. In its 2023 report \cite{IRI-ABA-Report}, the IRI task force identified several templates for interacting with HPC resources. Among these, data streaming offers powerful emerging capabilities such as near real-time data analysis, experimental steering, and informed decision-making during experiment execution. Unlike traditional store-and-forward models, which involve writing data to disk, transferring it over wide-area filesystems, and then reading it on a supercomputer, data streaming enables direct memory transfers. With direct memory streaming, bytes are moved straight from an edge node’s DRAM into an HPC job’s address space, bypassing intermediate storage layers. 

Facilities are actively exploring diverse technological solutions to enable direct, cross-facility data streaming. The benefits of evaluating different streaming architectures are multifold. First, performance assurance is critical, as streaming detectors or AI pipelines can burst to multi-gigabit rates, and each architectural hop introduces latency and jitter that impact performance under experimental load. In cross-facility scenarios, where institutions often operate within different security domains, controlled evaluations help quantify the impact of architectural choices on Transport Layer Security (TLS) overhead and trust boundaries. These comparisons also inform policy and compliance posture by guiding the redesign of institutional policies to support seamless integration across wide-area networks, coordinated resource provisioning, and guaranteed availability during experiments. Moreover, evaluating different architectures exposes trade-offs between control and abstraction, some offer fine-grained traffic and system tuning, while others emphasize managed, reproducible deployments. As facilities increasingly operate in hybrid environments that combine on-premise compute, cloud storage, edge processing, and user-deployed services, evaluations help identify architectures best suited for elastic scale and emerging experimental AI-HPC workflows. 

In this paper, we explore three architectures for cross-facility data streaming: Direct Streaming (DTS), Proxied Streaming (PRS), and Managed Service Streaming (MSS). These architectures are distinguished by how data flows from external producers (e.g., scientific instruments) to consumers (e.g., compute nodes in HPC jobs). All three rely on a dedicated backend streaming service to handle low-level mechanics such as connection management, buffering, and flow control. In DTS, data flows directly through exposed node-level network ports. PRS introduces intermediary proxies to relay data between facilities, while MSS routes data through facility-managed services using web-style domain names. By exploring these architectures, we aim to assess their network complexity, data flow paths, scalability, operational burden, performance characteristics, and overall feasibility for deployment and management in scientific workflows. Through simulated data streaming experiments using AI-HPC application communication patterns and workloads that closely mimic IRI science workflows, we study the streaming behavior of these architectures on a production-grade streaming infrastructure deployed at OLCF. 

Our paper makes the following contributions: 

\begin{itemize}
    \item We investigate the architectural differences and deployment models of three cross-facility data streaming architectures, DTS, PRS, and MSS, by incorporating a data streaming framework, DS2HPC, and a middleware toolkit, SciStream. 
    \item We characterize their streaming throughput, round-trip time, and overhead using three distinct workloads: two that closely emulate real IRI scientific workflows and one that represents a generic streaming scenario. 
    \item To conduct this study, we developed an in-house streaming simulator and evaluated the architectures using three types of messaging communication patterns: work sharing, work sharing with feedback, and broadcast and gather. 
\end{itemize}

%% file: tex/architecture.tex
\section{Cross-Facility Data Streaming Architectures}
\label{sec:architecture}

The choice of cross-facility data streaming architecture significantly impacts streaming performance, security posture, and deployment complexity. A comparative evaluation highlights trade-offs, helping facilities tailor deployments to their scientific and operational needs. This section examines the data flow from producers to consumers through a streaming service across three cross-facility architectures.

\subsection{Direct Streaming}
\begin{figure*}
\includegraphics[width=1\linewidth]{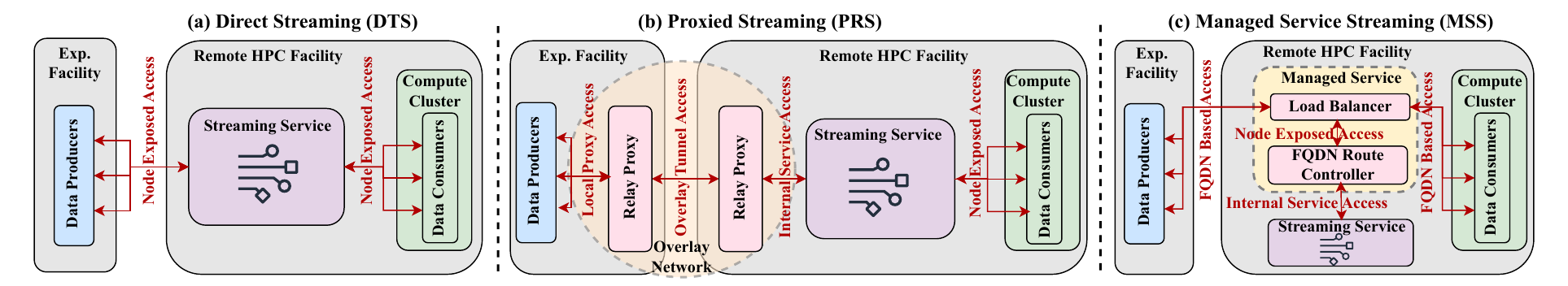}
\caption{\textbf{Three types of cross-facility data streaming architectures and their data flow paths.}}
\label{fig:architecture}
\vspace*{-3.5mm}
\end{figure*}

Direct Streaming (DTS) deploys the streaming service on destination nodes and exposes it via node-level network ports accessible to producers and consumers (see Figure \ref{fig:architecture}a). This minimal-hop path can reduce latency by eliminating intermediate proxies. However, enabling DTS requires explicit administrative configuration, including routing external traffic, opening firewall or iptables rules, and potentially assigning Domain Name System (DNS) entries. These steps are complicated by the presence of firewalls and Network Address Translation (NAT) between facilities, which typically block arbitrary access to internal nodes.

DTS introduces security risks by exposing node-level access and bypassing standard network policies. It scales poorly, as each new deployment demands manual port assignment and firewall updates. Despite offering a single-hop wide-area network (WAN) path via Destination NAT (DNAT), the operational burden of managing ports, certificates, and security makes it suitable mainly for tightly controlled environments. DTS is most viable within unified administrative domains, such as private research networks or peered subnets, where inter-site access is already trusted. For instance, OLCF's OpenShift \cite{openshift} clusters enable NodePort access via cluster-specific domains and ports, contingent on namespace policies \cite{olcf-nodeports}.

\subsection{Proxied Streaming}

Proxied Streaming (PRS) employs intermediary proxies to relay data between facilities, forming a logical overlay on top of physical infrastructure (see Figure \ref{fig:architecture}b). Rather than direct connections, producers and consumers communicate via local proxies, typically running on edge/gateway or Data Transfer Nodes (DTNs), which then forward data across the WAN through the overlay tunnel.

A key advantage of PRS is its ability to overcome NAT and firewall barriers with minimal, centralized firewall rules. By having one proxy initiate the connection, often through known gateway nodes, PRS avoids the complexity of managing node-level IPs and ports. Administrators can pre-authorize stable endpoints like DTNs, simplifying policy enforcement. PRS encapsulates traffic within a managed proxy layer that supports Mutual TLS (mTLS), authentication, and traffic control, enabling secure, scalable, and dynamically deployable streaming, suitable for cross-facility setups with proxies at DMZs \cite{dmz} or Virtual Private Network (VPN) endpoints.

SciStream \cite{scistream-paper, scistream-demo} is a toolkit based on PRS. Similar designs include TCP relays or message brokers on DTNs, service mesh gateways or VPNs in cloud environments, and Traversal Using Relays around NAT (TURN) servers in peer-to-peer systems, all of which act as overlay proxies when direct connectivity is unavailable.

\vspace*{-2mm}
\subsection{Managed Service Streaming}

In Managed Service Streaming (MSS), the facility's platform infrastructure manages the data flow. Producers send data to a stable Fully Qualified Domain Name (FQDN), which terminates at the HPC facility’s managed load balancer or ingress. A route controller then maps that hostname to the appropriate streaming endpoints (see Figure \ref{fig:architecture}c). The streaming service is provisioned on demand in response to user requests, and both producers and consumers connect via a standard DNS hostname and port (typically 443 for TLS), enabling seamless, web-style access.

This architecture abstracts away networking complexities: users do not manage IP addresses, ports, or NAT traversal. The facility handles all routing, DNS resolution, and security enforcement. Because the load balancer or gateway has a routable IP, only outbound connectivity is required at the producer site, simplifying firewall traversal. MSS offers high convenience, strong isolation, and %minimal configuration overhead. 
minimizes needs for configuration by the user.

%% file: tex/framework.tex
\section{Streaming Frameworks and Toolkits}
To deploy and test the streaming architectures, we use the OLCF’s Advanced Computing Ecosystem (ACE)~\cite{OLCF-ACE-FY24} infrastructure. The deployments leverage a data streaming framework, Data Streaming to HPC (DS2HPC), and a middleware toolkit, SciStream, both designed to address the challenges of cross-facility scientific data streaming.
\begin{figure*}
\centering
\includegraphics[width=0.8\linewidth]{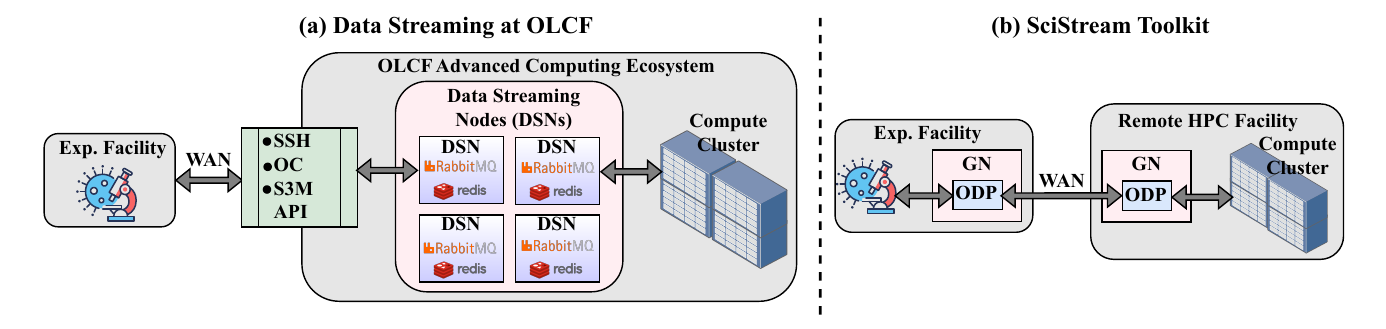}
\caption{\textbf{(a) Data streaming framework on ACE at OLCF, and (b) SciStream toolkit for memory to memory data streaming with on-demand proxies (ODPs) deployed on Gateway Nodes (GNs).}}
\label{fig:infrastructure}
\vspace*{-3mm}
\end{figure*}

\subsection{Data Streaming to HPC}
Data Streaming to HPC (DS2HPC) is an architectural framework for secure, bidirectional, memory-to-memory data streaming between science facilities and computing facilities~\cite{OLCF-DS2HPC}. As shown in Figure~\ref{fig:infrastructure}a, the OLCF has created an infrastructure within its ACE that supports demonstrations of the capabilities of the DS2HPC framework, as well as other streaming approaches, by installing Data Streaming Nodes (DSNs) as facility gateway nodes accessible via high-speed networks to both external and internal applications. The DSNs are managed as a unique namespace within an existing OpenShift cluster. DSNs can function either as application-aware gateways (OSI Layer 7) or as lightweight transport-layer routers (OSI Layer 4), depending on the deployment configuration. Access to deploy streaming technologies on DSNs is available via SSH, the OpenShift command-line (OC) interface, or through the Streaming API of the OLCF Secure Scientific Service Mesh (S3M)~\cite{skluzacek2025pearc}. 

S3M enforces a multi-layered security model based on time-limited, token-based authentication. At its core is Istio \cite{istio}, an open-source service mesh that provides fine-grained control over authentication, authorization, and policy compliance. Users are issued project-scoped tokens with well-defined permissions, and all incoming requests are validated against project allocations and policy rules. When a user submits a request via the Streaming API, S3M orchestrates the provisioning of the streaming service onto the requested number of DSNs. Current deployments support RabbitMQ \cite{RabbitMQ} and Redis \cite{Redis}, both optimized for high-throughput data paths. In addition to secure data streaming, S3M's Compute API supports dynamic compute orchestration, enabling users to trigger compute jobs on OLCF systems as part of their streaming workflows.

\vspace*{-3mm}
\subsection{SciStream}
SciStream is a memory-to-memory data streaming toolkit that enables data to flow directly from an instrument’s memory into remote HPC memory, even when neither system has direct network connectivity. It treats each experimental or HPC facility as an independent security domain and allows each to expose streaming resources through a federated interface. As shown in Figure~\ref{fig:infrastructure}b, the architecture hosts on-demand proxies (ODPs) on Gateway Nodes (GNs) within the facilities.

SciStream consists of three main components: the SciStream User Client (S2UC), which brokers requests and gathers short-lived credentials; the SciStream Control Server (S2CS), which runs on each GN and allocates local resources; and the SciStream Data Server (S2DS), a proxy that bridges the internal network and the WAN, authenticating external peers using proxy certificates and internal peers by source address.

A streaming session begins when the user’s S2UC contacts the producer-side and consumer-side S2CS instances to negotiate the number of parallel channels and the aggregate bandwidth available. Once accepted, a control protocol launches S2DS instances, assigns ports, builds a connection map, and signals the applications to begin transmission. The data then flow through three transparent hops: producer $\rightarrow$ local proxy $\rightarrow$ remote proxy $\rightarrow$ consumer, that are transparent to the applications. 

%% file: tex/deployment.tex
\vspace*{-2.3mm}
\section{Streaming Deployments}

We describe the deployment of all three streaming architectures on the ACE infrastructure utilizing both DS2HPC and SciStream. Here, the deployments use production-grade infrastructure; however, both producers and consumers reside within the same HPC cluster, thus emulating, rather than fully implementing, cross-facility streaming scenarios over a WAN.

\vspace*{-2mm}
\subsection{Olivine OpenShift Cluster}
Within the ACE infrastructure, the Olivine OpenShift cluster provides a suitable deployment platform, as it includes DSNs, dedicated gateway hosts that bridge the public WAN and internal OLCF networks via high-speed network adapters. Purpose-built for streaming, each DSN is equipped with two 32-core 2.70 GHz AMD EPYC 9334 processors and 512~GiB of RAM. While each node supports 100~Gbps connectivity to both internal and external networks, current usage is limited to 1~Gbps due to ongoing efforts to fully configure high-speed interfaces within the OpenShift environment. To deploy both producers and consumers we utilize OLCF's Andes \cite{andes} cluster.

\vspace*{-2mm}
\subsection{Streaming Service}
Off-the-shelf messaging frameworks are increasingly considered viable solutions for enabling direct memory-to-memory streaming, owing to their maturity, widespread adoption, and ability to abstract core messaging and reliability functions away from the application layer. In this study, we use the RabbitMQ messaging framework as the streaming service, chosen for its previous usage in prior scientific initiatives such as INTERSECT~\cite{intersect-toolkit, intersect-paper2}. RabbitMQ is a message broker that implements the Advanced Message Queuing Protocol (AMQP)~\cite{amqp} as its wire-level protocol. It allows producers to send messages, consumers to receive them, and queues to temporarily store messages until they are consumed. For our deployment, a three-node RabbitMQ (server version 4.0.5) cluster was deployed on the DSNs.

\subsection{DTS Deployment}
To deploy and configure a three-server RabbitMQ service for the DTS architecture, we used the Bitnami RabbitMQ Helm Chart (version v3.18.3) \cite{bitnami}. This chart provisions a RabbitMQ cluster with servers as separate pods on OpenShift, with specific security, performance, and networking configurations.

\begin{figure*}
\includegraphics[width=1\linewidth]{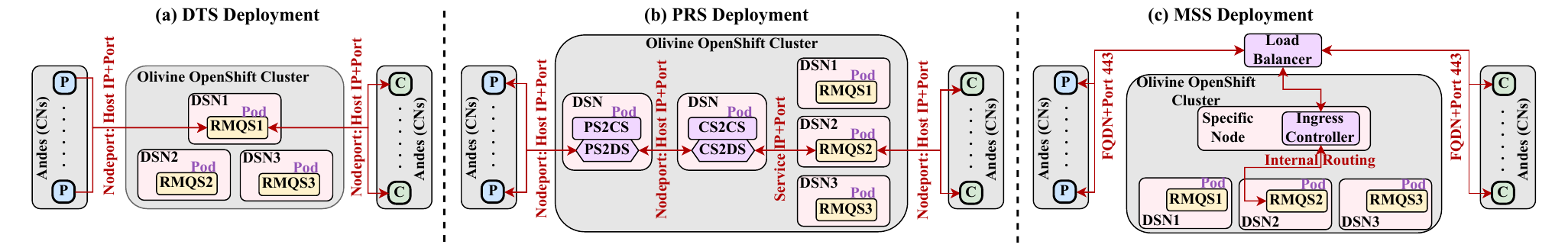}
\caption{\textbf{Deployments of DTS, PRS, and MSS architectures utilizing a three-server RabbitMQ cluster (RMQS1, RMQS2, and RMQS3) on the Olivine OpenShift cluster, with producers (P) and consumers (C) deployed on Andes compute nodes (CNs) at OLCF. In the PRS deployment, PS2CS and PS2DS represent the producer-side control server and data server SciStream components, while CS2CS and CS2DS represent the consumer-side components.}}
\label{fig:deployment}
\vspace*{-2mm}
\end{figure*}

The deployment enables TLS using auto-generated certificates, and each of the three server pods is scheduled on a separate DSN using pod anti-affinity rules. Resource requests are set to 12 CPUs and 32~GiB of memory per pod, with each replica allocated 15~GiB of persistent storage. The service is exposed via a NodePort (within the 30000-32767 range), with ports 30672 (AMQP) and 30671 (AMQPS) opened on the corresponding nodes. These node-exposed ports allow external access from both producers and consumers (see Figure \ref{fig:deployment}a). While DTS is feasible only between sites with direct connectivity, evaluating it helps quantify the streaming overhead of the other architectures in comparison.

The RabbitMQ deployment command using Helm Chart is given below in which rabbit.yaml is the file where cluster configurations are specified.
\framedtext{\textcolor{blue}{helm install rabbitmq bitnami/rabbitmq --namespace abc123 -f rabbit.yaml}}

%\vspace*{-2mm}
\subsection{PRS Deployment}
In the PRS setup using SciStream the producer and consumer S2CS instances were deployed as OpenShift pods on two separate DSNs (see Figure \ref{fig:deployment}b). For establishing the overlay tunnel, SciStream supports three options: Stunnel, HAProxy, and Nginx. In this deployment, we evaluated two tunneling methods - Stunnel and HAProxy. Additionally, SciStream offers flexibility to increase the number of parallel connections from producers and consumers to their respective proxies. We used SciStream version 1.2.1 \cite{scistream-github} for this setup. Upon startup, the container process in the S2CS pod determines its own IP address, generates a self-signed TLS certificate using openssl \cite{openssl}, and launches the S2CS process with TLS enabled. The container exposes ports 5000 and 5100–5110 for streaming traffic. 

Note that the PRS architecture (Figure \ref{fig:architecture}b) shows local proxy access and overlay tunnel connections---specifically in the first two hops. However, for the proof-of-concept deployment, we exposed each hop using \texttt{<node-IP>:<NodePort>} to minimize external dependencies and decouple the overlay logic from load balancer or ingress configurations.

Each S2CS pod is paired with a NodePort service that exposes the pod's streaming ports for node-level access. These nodePorts are used by the proxies launched by the S2CS instances to facilitate node-exposed access. For the streaming service itself, we use the same Helm chart-based three-node RabbitMQ cluster described in the DTS setup. The only difference is in protocol usage: in DTS, TLS-encrypted (AMQPS) data is used directly; however, in the PRS setup, we use AMQP without TLS, since SciStream’s overlay tunnel already provides encryption and authentication via its TLS layer. In a true cross-facility deployment, the only path exposed externally is the overlay tunnel, meaning all data traffic traversing this route is protected by SciStream’s TLS-secured tunnel. Once the data reaches the RabbitMQ server inside the HPC facility, it is already encrypted, and thus additional TLS between the consumer S2CS and RabbitMQ is unnecessary. 

The S2UC was deployed as an Apptainer \cite{apptainer} container on a login node in the Andes cluster. S2UC issues two requests to create the respective proxies (S2DSs) and establishes the overlay tunnel between them---an inbound request and an outbound request.
The commands for inbound and outbound requests are as follows:

\framedtext{\textcolor{blue}{s2uc inbound-request --server\_cert=../certs/cons-s2cs.crt --remote\_ip 10.1.1.100 --s2cs 198.51.100.0:30600 --receiver\_ports 5672 --num\_conn 1}}
An output of the inbound request would be an S2DS consumer side proxy (PROXY) and a unique identifier (UID) for the request, both of which are used in the outbound request command send to the producer to create producer proxy.
\framedtext{\textcolor{blue}{s2uc outbound-request --server\_cert=../certs/prod-s2cs.crt --remote\_ip 198.51.100.0 --s2cs 198.51.100.1:30500 --receiver\_ports PROXY --num\_conn 1 UID 198.51.100.0:PROXY}}
An output of the outbound proxy command would be an S2DS producer side proxy.

\subsection{MSS Deployment}
In the facility-managed MSS deployment, as shown in Figure \ref{fig:deployment}c, the load balancer is dedicated hardware located outside the OpenShift cluster. It forwards traffic to the cluster’s OpenShift ingress controller, a core platform component that runs on separate ingress nodes rather than on the DSNs. A three-node RabbitMQ cluster, configured with the same specifications as in the other two deployments, is provisioned via the facility’s S3M API \cite{s3m-docs}.

The command to provision the RabbitMQ cluster is shown below. It requires an authentication token generated via the S3M API and allows specifying the cluster’s resource requirements. Upon execution, the command returns an FQDN-based AMQPS URL that users can directly provide to their RabbitMQ client connection API.
\framedtext{
\textcolor{blue}{curl -X POST ``https://s3m.apps.olivine.ccs.ornl.gov/olcf/v1alpha/ streaming/rabbitmq/provision\_cluster'' -H ``Authorization: TOKEN'' -H ``Content-Type: application/json'' -d `{``kind'': ``general'', ``name'': ``rabbitmq'', ``resourceSettings'':{``cpus'': 12, ``ram-gbs'': 32, ``nodes'': 3, ``max-msg-size'': 536870912}}'}}

%% file: tex/evaluation.tex
\section{Evaluation}
We conducted an emulation-based evaluation of cross-facility streaming architectures by deploying real messaging components within a single HPC cluster. While not a true multi-site setup, this environment preserves the key network characteristics and data flow paths described in \S\ref{sec:architecture}.

\vspace*{-2mm}
\subsection{Messaging Patterns and Streaming Workloads}
\label{sec:params_workloads}
To conduct the evaluation, we selected three messaging patterns: \textbf{work sharing, work sharing with feedback, and broadcast and gather}. These patterns align closely with communication motifs found in many AI and HPC workloads. The work sharing pattern is used in embarrassingly parallel workloads such as hyperparameter searches \cite{hyperparams} or Monte Carlo ensembles \cite{monte-carlo}, where thousands of short, independent jobs are launched without any post-dispatch communication. Schedulers like Slurm job arrays or GNU Parallel distribute tasks once, allowing each node to compute in isolation, matching the work sharing model.

% TODO: cite Junqi / Feiyi / Wes Brewer's work on AI model serving - already cited junqi's neutron paper and Wes' AI models paper (Anjus)
The work sharing with feedback pattern appears in data-parallel deep learning (DL) frameworks like TensorFlow-PS \cite{tensorflow} and MXNet \cite{mxnet}, where a weight shard is sent to each worker, and each worker returns its gradient to the same shard. This push–pull interaction forms a classic distribute-with-reply loop. Another example is master–worker task farms that require immediate results, such as real-time inference on micro-batches, where each reply must be routed back to the originating producer. The broadcast and gather pattern is common in distributed data-parallel (DDP) training frameworks like NCCL \cite{nccl} and Gloo \cite{gloo}, which perform a fan-out of weights, a reduce-scatter of gradients, and an all-gather of aggregated values every iteration. Another instance is large-scale metric aggregation, where a single node issues a collect request and all workers send back metrics to be reduced at the initiator.

We now define three streaming workloads: two based on IRI science workflows and one representing a generic streaming scenario. The IRI workflows were selected from the OLCF Science Pilots and Workflows initiative, which aims to implement them on the ACE infrastructure to support experimental steering and cross-facility integration across diverse scientific domains \cite{OLCF-ACE-FY24}. Two representative workflows, GRETA/Deleria \cite{greta-deleria} and SLAC-LCLS \cite{lcls}, serve as strong use cases for cross-facility data streaming.

GRETA (Gamma-Ray Energy Tracking Array) is a gamma-ray spectrometer currently being deployed at the Facility for Rare Isotope Beams at Michigan State University. It enables real-time analysis of gamma-ray energy and 3D position with up to 100x greater sensitivity than existing detectors. The associated workflow software, Deleria, continuously streams experimental data over ESNet to hundreds of analysis processes on an HPC system, processing up to 500K events per second. Deleria supports time-sensitive streaming and has been deployed across ESNet and ACE to demonstrate a distributed experimental pipeline. Recent emulation experiments on ACE scaled to 120 simulated detectors, achieving sustained bi-directional streaming rates of $\sim$35~Gb/s.  

The Linac Coherent Light Source (LCLS) at SLAC National Accelerator Laboratory provides X-ray scattering for molecular structure analysis and streams experimental data to enable rapid analysis and decision-making between experiment runs. The LCLStream pilot project trains a generalist AI model using streamed detector data, from both archived and live LCLS/LCLS-II \cite{lcls-2} experiments, to support tasks like hit classification, Bragg peak segmentation, and image reconstruction. This AI-driven approach serves as a shared backbone for various downstream data analysis tasks. With the new LCLS-II producing data at 400x the rate of its predecessor, streaming up to 100~GB/s to HPC systems will be essential for responsive analysis and experiment steering. LCLStream aims to support online streaming and real-time analysis during experiment execution, eliminating delays associated with waiting for data to be written to file storage systems before processing. 

\begin{table}[h!]
  \caption{Data streaming characteristics for streaming workloads - Deleria, LCLS and generic workloads.}
  \label{tab:workflows}
  \vspace*{-3mm}
  \fontsize{7pt}{7pt}\selectfont
  \begin{tabular}{|>{\raggedright\arraybackslash}p{0.6in}|>{\raggedright\arraybackslash}p{0.7in}|>{\raggedright\arraybackslash}p{0.7in}|>{\raggedright\arraybackslash}p{0.7in}|}
    \hline
    \textbf{Characteristics} & \textbf{Deleria} & \textbf{LCLS} & \textbf{Generic} \\
    \hline
    Payload size & $\approx$KiB range & $\approx$1 MiB & 4 MiB \\
    \hline
    Payload format & Binary & HDF5 & Binary \\
    \hline
    Payload element & Events & Events & Variables \\
    \hline
    Data packaging & Variable \# events/msg  & Variable \# events/msg & One item/msg\\
    \hline
    Data rate & 32 Gbps & 30 Gbps & 25 Gbps \\
    \hline
    Consumption parallelism & Parallel (non-MPI) & Parallel (MPI-based) & Parallel (MPI-based) \\
    \hline
    Production parallelism & Parallel (non-MPI) & Parallel (MPI-based) & Parallel (MPI-based) \\
    \hline
  \end{tabular}
  \vspace*{-3mm}
\end{table}

Table \ref{tab:workflows} shows the key data streaming characteristics relevant to all the workloads. The LCLS stream uses $\approx$1 MiB data payloads with a steady data rate of $\approx$30 Gbps sustained over 1–100 minutes. Each message contains an HDF5-formatted file, with producers and consumers launched using MPI. Messages are pushed to consumers in a round-robin fashion as they become available in the queue.  

In contrast, Deleria streams messages in the KiB range, each containing multiple experimental events batched together. The number of events per message is variable. Data messages use a compressed binary format, while control messages are encoded in JSON. Depending on the type of experiment, the GRETA detector sustains a steady data rate of up to 32~Gbps once an experiment begins. Producers and consumers do not use MPI. Instead, consumers pull event batches asynchronously from a remote forward buffer, while pushing processed events to a remote event builder. Although the Deleria workload’s payload size is variable in the KiB range and streams a variable number of events per message, for consistency, we fix the payload size to 2~KiB per event and the number of events per message to eight, resulting in a 16~KiB message size. 

The third workload is a generic scenario with arbitrarily defined streaming characteristics. We use the Deleria and LCLS streaming workloads, referred to as \textbf{Dstream} and \textbf{Lstream}, respectively, to evaluate the first two communication patterns: work sharing and work sharing with feedback. The \textbf{generic} workload is used to evaluate the third pattern, broadcast and gather.

\subsection{Simulator and Messaging Parameters}
To simulate the streaming experiments, we developed a Golang-based simulator\footnote{Our data streaming simulator and configuration files are publicly available here: \url{https://github.com/Ann-Geo/StreamSim/}} and utilized the amqp091-go (version 1.10.0) \cite{amqp-go} RabbitMQ AMQP client library to implement RabbitMQ APIs. The simulator accepts the streaming characteristics of workflows, as listed in Table \ref{tab:workflows}. Additionally, the simulator allows specifying type of streaming architecture (e.g., DTS, PRS, or MSS), streaming service specific parameters (e.g., type of acknowledgements, number of queues, prefetch count), experiment configurations (e.g., number of producers and consumers, message count, experiment duration), and infrastructure or toolkit specific options (e.g., URL for connection, number of connections, TLS). For a given message count or test duration, the simulator runs the experiment with the specified number of producers and consumers. Each producer is identical in function and is responsible for generating workload based on the input workload characteristics and sending data to the RabbitMQ server according to the specified parameters. Similarly, each consumer is identical and is designed to receive messages from the RabbitMQ server based on the same set of parameters.  

In addition to the producers and consumers, the simulator includes a coordinator component that serves two primary functions. First, it informs producers and consumers about which queues to use. Second, it collects metrics from individual consumers/producers and reports the aggregate results for the entire experiment. Each component, upon startup, handles the initialization of all required queues for the experiment run. The simulator supports launching both MPI-based and non-MPI producers and consumers. For the simulator clients, a total of 33 nodes from OLCF’s Andes system \cite{andes} were used: 16 nodes for producers, 16 nodes for consumers, and one node for the coordinator. Each Andes node is equipped with two 16-core 3.0~GHz AMD EPYC 7302 processors and 256~GiB of RAM. Andes and the DSNs in Olivine cluster are connected via a 1~Gbps Ethernet network.  

For the simulations presented, we measured two metrics: throughput and round-trip time (RTT). Throughput refers to the aggregate message rate (messages per second) from all consumers involved in each experiment. RTT is the time it takes for a message to travel from a producer to a consumer and for the corresponding reply to return to the producer. Additionally, from the measured metrics, we calculate the streaming overhead of the other two architectures relative to DTS, since DTS serves as a baseline with direct connectivity and no intermediate proxies. Each data point represents the average of three runs, with each run streaming up to 128K messages. Except for the broadcast and gather pattern, which uses one producer and many consumers, all other tests were performed with an equal number of producers and consumers to evaluate scaling behavior. In all experiments, consumers were started before producers.

To align streaming behavior with the workload characteristics shown in Table~\ref{tab:workflows}, we configure RabbitMQ with specific parameters. For the Dstream and Lstream workloads using the work sharing pattern, we adopt the work queue model, where producers send messages to shared queues and messages are distributed nearly evenly across multiple consumers. For the work sharing with feedback pattern, we use the same work queue model for request messages, but employ the direct routing model for replies. Each producer has a dedicated reply queue, ensuring that replies are routed back to the correct producer. This prevents misrouting and eliminates the risk of a producer waiting indefinitely for a reply intended for it but consumed by another. For both the above patterns we used two shared work queues to achieve increased throughput \cite{stream-study}.

In the broadcast and gather pattern, we use the publish-subscribe (pub-sub) model for both requests and replies. Requests are broadcast to all consumers, and the replies are sent to another pub-sub queue from which the single producer consumes all responses. All queue models use RabbitMQ’s classic queues, which retain a fixed number of messages in memory and support configurable durability. We set the queue overflow policy to ``reject-publish'', allowing producers to detect backpressure, handle rejected messages, and attempt republishing. Of the total RAM allocated to RabbitMQ servers, 80\% is reserved for data payload queues, with the remaining 20\% allocated to additional queues used for control messages, and workflow simulation management. Additionally we enable batch wise producer and consumer acknowledgements for guaranteed message reception. We now present the results and insights from our evaluation for all patterns.

\subsection{Work Sharing Pattern}

\begin{figure}
\includegraphics[width=\linewidth]{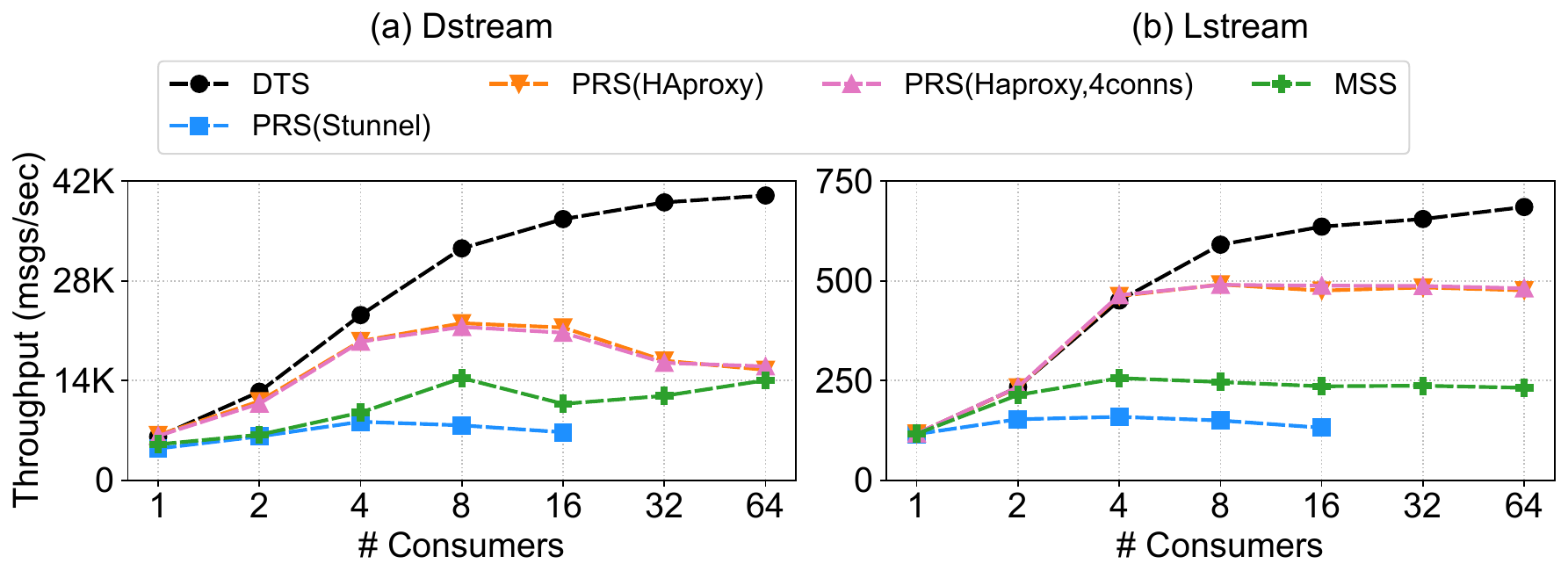}
\vspace*{-5mm}
\caption{\textbf{Throughput (msgs/sec) for (a) Dstream and (b) Lstream workloads under work sharing pattern.}}
\label{fig:del-lcls-throughput}
\vspace*{-6mm}
\end{figure}

Figure \ref{fig:del-lcls-throughput} shows the aggregate throughput for DTS, PRS, and MSS architectures. For the PRS setup, we also evaluated two additional SciStream tuning options: network proxy type and number of connections to proxies. Specifically, we tested Stunnel and HAProxy proxies, with up to four connections used in the HAProxy configuration. For the Dstream workload with a single producer and consumer, PRS with HAProxy achieved the highest throughput at 6.3K msgs/sec, while other configurations ranged between 4.4K and 6.2K msgs/sec. As the number of producers and consumers increased, DTS throughput improved, reaching a maximum of 39K msgs/sec at 64 consumers, though signs of saturation were observed beyond 32 consumers.

In the PRS configuration using Stunnel, throughput showed no significant improvement beyond a single consumer. Moreover, Stunnel could support a maximum of 16 simultaneous connections in our setup, making configurations with 32 and 64 consumers infeasible (no data points shown). This limitation is due to Stunnel’s design, which favors a few long-lived, TLS-wrapped flows rather than load balancing. By contrast, PRS with HAProxy scaled better, reaching up to 19K msgs/sec with four consumers, but throughput stagnated and began to decline beyond eight consumers. Increasing proxy connections to four showed no significant performance gain. The MSS architecture achieved slightly lower throughput than PRS with HAProxy, but consistently outperformed PRS with Stunnel. However, MSS also showed scaling limitations beyond eight consumers, with a maximum throughput of 14K msgs/sec.

For the Lstream workload, which uses a larger payload size, DTS achieved a maximum throughput of 685 msgs/sec at 64 consumers, but saturation occurred beyond eight consumers. PRS with HAProxy scaled well up to four consumers, after which throughput plateaued. PRS with Stunnel showed similar limitations as with Dstream. MSS throughput saturated beyond four consumers, peaking at 256 msgs/sec. Overall, in the work-sharing pattern experiments, we observed significant overhead in the PRS and MSS setups (up to 2.5x) compared to DTS, due to the presence of proxies in PRS and the load balancing and ingress mechanisms in the MSS architecture.

\begin{figure*}[t]
\includegraphics[width=1\linewidth]{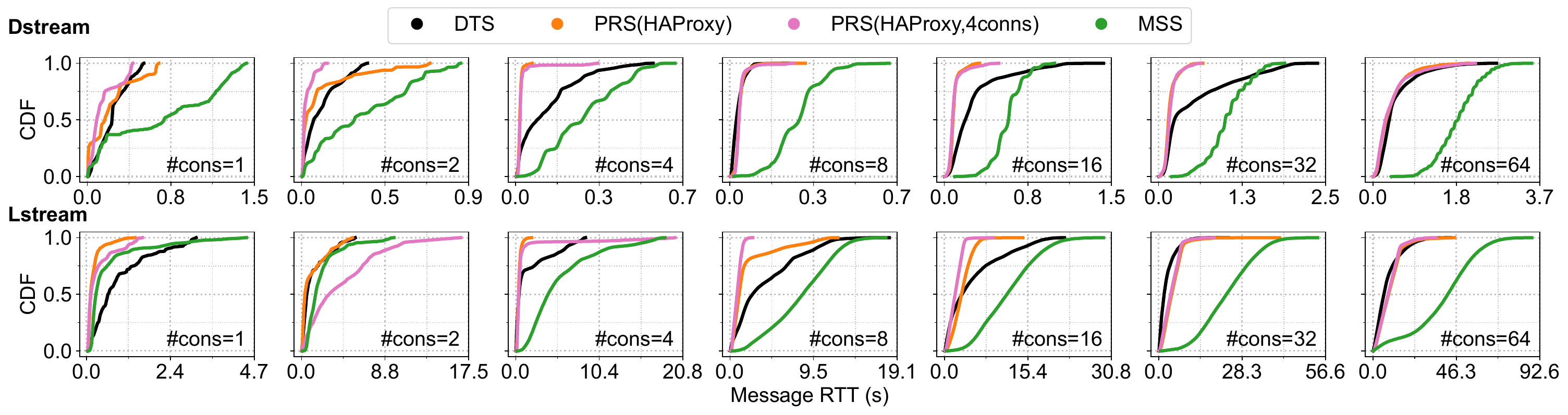}
\caption{\textbf{CDF of individual message RTTs for Dstream and Lstream workloads under work sharing with feedback pattern, with the number of consumers varying from 1 to 64.}}
\label{fig:del-lcls-latency-cdf}
\vspace*{-5mm}
\end{figure*}

\subsection{Work Sharing with Feedback Pattern}
\begin{figure}[h!]
\includegraphics[width=\linewidth]{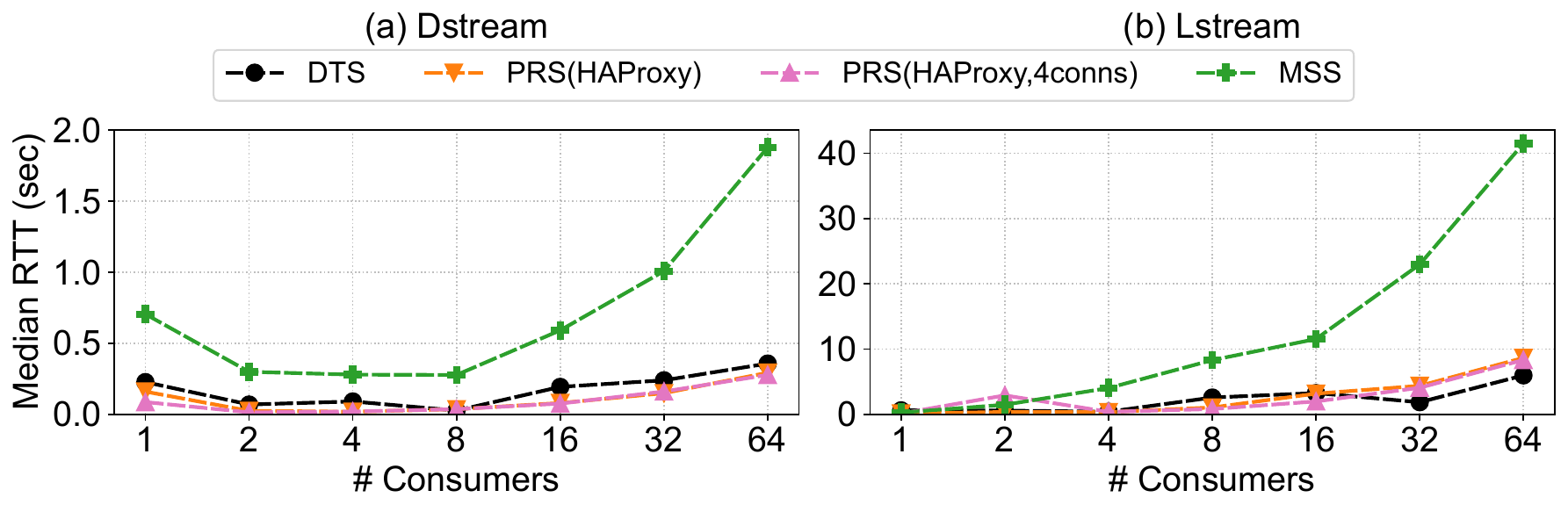}
\caption{\textbf{Median RTT (sec) for (a) Dstream and (b) Lstream workloads under work sharing with feedback pattern.}}
\label{fig:del-lcls-median-rtt}
\vspace*{-5mm}
\end{figure}

In the work-sharing with feedback pattern, we measured the per-message round-trip time (RTT), the time taken for a message to travel from a producer to a consumer and return to the producer. Figure~\ref{fig:del-lcls-median-rtt} shows the median RTT for all three architectures. Since PRS with Stunnel showed poor performance in earlier tests, we excluded it from further RTT evaluations and focused only on PRS with HAProxy.

For Dstream workload, both PRS and DTS, maintained median RTTs under 0.5 seconds across all consumer counts. DTS recorded a minimum RTT of 20ms, while PRS with HAProxy achieved a minimum of 17ms with eight and four consumers, respectively. RTT increased slightly for both architectures as the number of consumers grew beyond eight. The MSS setup consistently showed higher RTTs and experienced a sharp increase at 64 consumers, peaking at 1.8 seconds.

In the Lstream workload, for a single consumer, RTTs were under 200ms for both DTS and PRS, but 600ms (3x higher) for MSS. Median RTTs for DTS and PRS remained below 10 seconds across all consumer counts. In contrast, the MSS setup showed a significant RTT spike beyond eight consumers, with a severe bottleneck at 64 consumers, where the median RTT reached 40 seconds. 

In contrast to the work-sharing pattern, where PRS showed significant overhead, the work-sharing with feedback pattern demonstrated similar or slightly better performance for PRS compared to DTS. However, MSS continued to exhibit significant overhead (6.9x) in this pattern as well, with RTTs increasing sharply.

Since median latency cannot capture per-message RTT variations, we present the CDF of RTTs for all messages in Figure \ref{fig:del-lcls-latency-cdf}. Beyond eight consumers, all architectures exhibit a noticeable rightward shift in the CDF, particularly in the MSS architecture under the Lstream workload. This is likely due to the direct feedback path from consumers to producers, which introduces significant RTT bottlenecks. In contrast, the PRS architecture consistently maintains tighter RTT distributions, with less variation than MSS, and often performs comparably to DTS. Notably, for the 64-consumer case, PRS keeps 80\% of message RTTs under 0.7 seconds for Dstream and 12.5 seconds for Lstream, demonstrating uniformity in latency. However, increasing PRS connections from one to four does not yield observable improvements in RTT. 

\subsection{Broadcast and Gather Pattern}
\begin{figure}
\includegraphics[width=\linewidth]{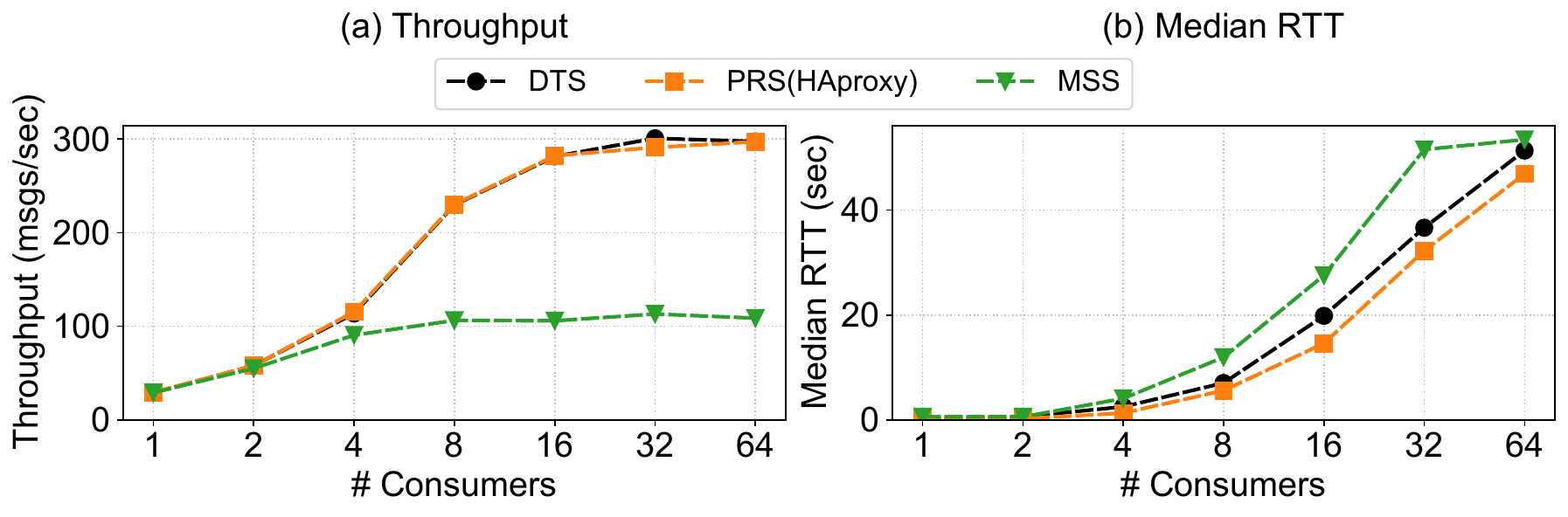}
\caption{\textbf{(a) Throughput (msgs/sec) and (b) Median RTT for generic workload under broadcast, and broadcast and gather patterns respectively.}}
\label{fig:generic-through-rtt}
\vspace*{-5mm}
\end{figure}

As mentioned in Section \ref{sec:params_workloads} we use the generic workload defined in Table \ref{tab:workflows} for these experiments. In the first experiment (Figure \ref{fig:generic-through-rtt}a), a single producer broadcasts the same message to all consumers, and we measure the aggregate consumer throughput. We observe that the PRS architecture scales almost equivalently to the DTS architecture. In contrast, MSS experiences bottlenecks beyond four consumers, with throughput stagnating around 110 msgs/sec. DTS and PRS also exhibit stagnation, but only beyond 32 consumers. Because the generic workload uses a larger payload size of 4~MiB, the 1~Gbps network link between consumers and the streaming service can saturate quickly. 

\begin{figure*}[t]
\includegraphics[width=1\linewidth]{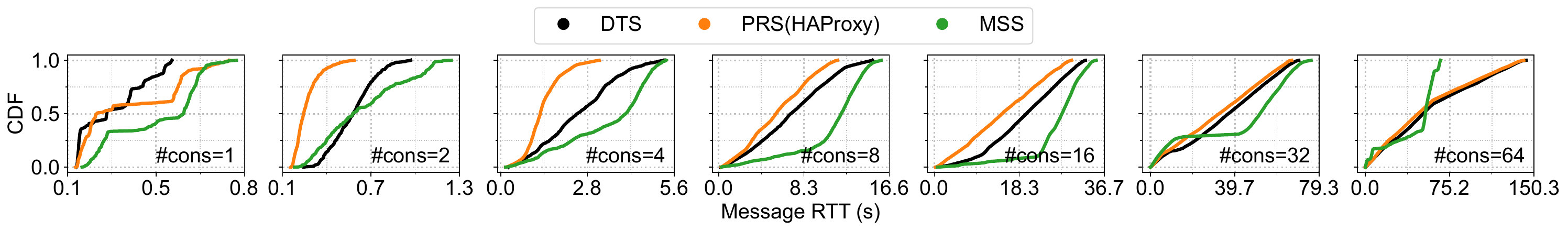}
\caption{\textbf{CDF of individual message RTTs for generic workload under broadcast and gather pattern, with the number of consumers varying from 1 to 64.}}
\label{fig:generic-rtt-cdf}
\vspace*{-5mm}
\end{figure*}

Figure \ref{fig:generic-through-rtt}b shows the second experiment, where, along with the broadcast described above, the producer also gathers replies from all consumers. In this case, we measure the median RTT. Unlike the work sharing with feedback pattern, where DTS and PRS had similar median RTTs and MSS differed significantly, here we observe that all three architectures exhibit comparable median RTTs. For up to four consumers, all RTTs are under five seconds. Beyond this point, RTTs increase sharply due to a single-producer bottleneck, where the same producer must handle both broadcast and gather operations for all consumers. 

The CDF trends of per-message RTT in this pattern (Figure \ref{fig:generic-rtt-cdf}) show that, in some cases, particularly with 2 to 16 consumers, PRS exhibits lower RTTs than DTS. MSS still shows comparatively higher RTTs, though not as high as in the work sharing with feedback pattern. Another notable trend is that as consumer count reaches or exceeds 32, the CDFs of all three architectures converge, and MSS begins to outperform the other two.

%% file: tex/discussion.tex
\section{Conclusion, Discussion, and Future Work}

In this paper, we investigated the architecture, deployment, and performance of three cross-facility data streaming architectures - DTS, PRS, and MSS, using workloads derived from experimental-HPC scientific workflows and AI-HPC application communication patterns. Architecturally, DTS offers the lowest-hop path but requires explicit firewall/iptables rule configurations. PRS reduces such requirements via stable, pre-authorized endpoints, and MSS fully abstracts networking by letting facility's platform manage data flow. Deployment-wise, DTS is simplest but requires manual configurations, PRS adds moderate complexity with proxy/tunnel setup, and MSS is the most streamlined through managed APIs. In the evaluation, PRS and MSS experience significant overhead in the work sharing pattern. In the work sharing with feedback pattern, PRS performs as well as or better than DTS, while MSS continues to show overhead. In the broadcast and gather pattern, PRS scales almost equivalently to DTS for broadcast throughput, and in the gather phase, as consumers scale up, all architectures converge, with MSS eventually showing lower latency trends.

% TODO: reference INTERSECT projects - already did (Anjus)

%The trade-off for application layer awareness is the potential for increased latency and lower message throughput.
%Comparing direct, proxied, and managed data transmission architectures shows the expected trend that increasing the number of network hops lowers overall performance.  Both metrics can increase, however, for architectures that can make use of parallel message streams.  

\textbf{Other streaming architectures}: In future, we plan to compare these results with streaming architectures that operate at the network layer, with potentially reduced guarantees on message delivery. The EJFAT project \cite{ejfat} has produced an FPGA-accelerated router gathering data in the form of UDP network packets from multiple producers and forwarding it to multiple consumers.  Internal to OLCF, there is an effort called Project Banana Pepper to configure network routers as NAT gateways, selectively forwarding traffic associated with a specific set of compute nodes.  Early results from those works show that network-layer forwarding is feasible and maintains high-level control over network reachability.

\textbf{Usage of high-speed network}: The evaluation shows that throughput eventually stagnates for all architectures as the number of consumers increases. A primary reason is the 1~Gbps network link between the compute nodes (producers/consumers) and the DSNs (RabbitMQ servers). While the DSNs have 100~Gbps interfaces, configuration issues within our OpenShift environment have prevented their effective use. The transition of the OS from RedHat Enterprise Linux (RHEL) to RedHat CoreOS (RHCOS) added several new technical challenges. RHCOS is an immutable operating system where the OS root filesystem image is put into a standard Open Container Initiative (OCI) container, where the OS base image is build up with layers into a Dockerfile. Partitioning the 100~Gbps interfaces with Single Root I/O Virtualization (SRIOV) using RHCOS layering has resulted in degraded virtual interfaces that provide only marginal bandwidth ($<$1~Gbps). Further work is needed to make the 100~Gbps interfaces fully utilizable. However, this does not affect our comparative results, since we included the DTS architecture, which has direct connectivity and no intermediate proxies, as a baseline. Although DTS is only feasible in environments with direct connectivity, it provides a reference point for quantifying the overhead introduced by the other two architectures.

\textbf{Challenges in making PRS (SciStream) usable for external users}: SciStream requires clients to connect to a reachable IP and port. However, external access in the OLCF OpenShift cluster is restricted to HTTPS routes over port 443. Because SciStream uses custom ports, additional firewall rules or policies may be needed to permit external traffic. At OLCF, routing on port 443 is hostname-based, whereas SciStream identifies connections by port numbers or unique identifiers (UIDs). For compatibility, SciStream would need new features to support or adapt to hostname-based routing.

\textbf{Usage of a streaming service}: Application-layer message forwarding offers several advantages for designing robust data science pipelines. In work sharing mode with reliable message delivery, data receivers can make use of high degrees of parallelism while being assured that rare events will not be lost---a key need for GRETA/Deleria.  Forwarding at this layer can also enable feedback to data producers and broadcast-gather patterns. These are necessary in reliably coordinating experimental instruments and feedback loops with automated control systems.

\textbf{Network and gateway load balancers}: % I would defer to AWS / Azure here
As data handling requirements increase, message forwarding services will need to take advantage of load balancing.  This can introduce additional complexity into the application if it needs to be aware of multiple network addresses, or into the streaming framework if (as described in this work) a multi-address DNS scheme is used.  Our major conclusions hold about the trade-off between network-level and application level message forwarding.  However, cloud services have made several innovations in this space---especially in distributing HTTP traffic---building configurable load balancers to decrease the processing time of client requests.

\textbf{MSS scaling limitations and bottlenecks}: Our evaluation shows that MSS encounters scaling limits and severe bottlenecks in both throughput and latency when producers and consumers are scaled up. One possible improvement is to allow compute nodes inside the facility to bypass the load balancer and connect directly to streaming pods on the DSNs, reducing hops and potential contention. 